\documentclass{PoS}

\newcommand{\tmtexttt}[1]{{\ttfamily{#1}}}
\usepackage[square,sort,comma,numbers]{natbib}
\RequirePackage{nccmath}
\RequirePackage{enumerate} 
\RequirePackage{subcaption}
\RequirePackage{wrapfig} 
\usepackage{url}

\title{Second large-scale Monte Carlo study for the Cherenkov Telescope Array}

\ShortTitle{Second large-scale Monte Carlo study for the Cherenkov Telescope Array}

\author{\speaker{T. Hassan}$^{1}$, L. Arrabito$^{2}$, K. Bernl\"{o}hr$^{3}$, J. Bregeon$^{4}$, J. Hinton$^{3}$, T. Jogler$^{5}$, G. Maier$^{6}$, A. Moralejo$^{1}$, F. Di Pierro$^{7}$, M. Wood$^{5}$ for the CTA Consortium\footnote{Full consortium list at http://cta-observatory.org}\\
        $^{1}$IFAE, Edifici Cn., Campus UAB, E-08193 Bellaterra, Spain\\
        $^{2}$C-IN2P3 - Centre de Calcul (FR)\\
        $^{3}$Max-Planck-Institut f\"{u}r Kernphysik\\
        $^{4}$Laboratoire Univers et Particules de Montpellier, Universit\'{e} Montpellier\\
        $^{5}$SLAC National Accelerator Laboratory, Stanford University\\
        $^{6}$Deutsches Elektronen-Synchrotron (DESY)\\
        $^{7}$Istituto Nazionale di Astrofisica (INAF)\\
        E-mail: \email{thassan@ifae.es}}

\abstract{\color{black}The Cherenkov Telescope Array (CTA) represents the next generation of ground based instruments for Very High Energy gamma-ray astronomy. It is expected to improve on the sensitivity of current instruments by an order of magnitude and provide energy coverage from 20 GeV to more than 200 TeV. In order to achieve these ambitious goals Monte Carlo (MC) simulations play a crucial role, guiding the design of CTA. Here, results of the second large-scale MC production are reported, providing a realistic estimation of feasible array candidates for both Northern and Sourthern Hemisphere sites performance, placing CTA capabilities into the context of the current generation of High Energy $\gamma$-ray detectors.}

\FullConference{The 34th International Cosmic Ray Conference,\\
		30 July- 6 August, 2015\\
		The Hague, The Netherlands}

\begin{document}

\section{Introduction}

As a result of the success of current Imaging Atmospheric Cherenkov Telescopes (IACTs) and the improvement of the different technologies involved, the next generation of ground-based Very High Energy (VHE) $\gamma$-ray detectors is under development. The Cherenkov Telescope Array (CTA)\footnote{\url{http://www.cta-observatory.org/}} \cite{CTA_concept} will scrutinize the $\gamma$-ray sky from some tens of GeV up to more than a hundred TeV with unprecedented capabilities, improving the sensitivity of current instruments by up to an order of magnitude. 


In order to achieve these goals, the CTA Observatory will consist of two different sites, one in each hemisphere, and telescopes of three different size classes: Large Size Telescopes (LSTs) sensitive to the faint-low energy showers (below 200 GeV),  Medium Size Telescopes (MSTs) increasing the effective area and the number of telescopes simultaneously observing each event (\textit{multiplicity}) within the CTA core energy range (between 100 GeV and 10 TeV) and Small Size Telescopes (SSTs, for CTA South only) spread out over several km$^2$ to catch the rare events at the highest energies of the electromagnetic spectrum (up to $\sim$ 300 TeV). The current design foresees 4 LSTs, 25 Davies Cotton (DC) MSTs, and 70 SSTs for CTA South as well as 4 LSTs and 15 MSTs for CTA North. The southern array may be augmented with a proposed extension of up to 25 Schwarzschild-Couder (SC) MSTs.

The proposed designs for the northern and southern observatories will provide full sky coverage with an improved sensitivity alongside better angular and energy reconstruction. The CTA southern site is expected to be larger, composed of $\sim 100$ telescopes, to take advantage of its privileged location to observe the Galactic Plane and Galactic Center over the full VHE range. The northern site is expected to complement it with a reduced number of telescopes of the order of $\sim 20$ (LSTs and MSTs).


In this work we present results from the second large-scale MC production (Prod-2), and provide a realistic estimation of the future observatory capabilities. Prod-2 results concerning the site selection (such as the performance dependencies on altitude or the local Geo-magnetic field) and the performance of mixed MST type layouts are presented as separate contributions \cite{MC_ICRC_site:2015, SCTs_ICRC:2015}.

\section{Monte Carlo simulations and analysis tools}
\label{sec:MCsim}

CTA Monte Carlo simulations, as detailed in \cite{APP_CTA_MC}, are performed by defining a large telescope layout, comprising few hundreds of telescopes of different types arrayed over an area of around 6 km$^2$. Simulations of extensive air showers (EAS) initiated by $\gamma$-rays, cosmic-ray nuclei and electrons are generated using CORSIKA \cite{corsika} together with the simulation of the telescope response using \tmtexttt{sim\_telarray} \cite{Konrad:2008}. By selecting sub-samples of telescopes of approximately equal costs from the macro-layout and evaluating their performance, conclusions on the most cost-effective CTA array are inferred. 
 
The first large-scale MC production (Prod-1) already simulated a wide variety of conditions (using a constant geomagnetic field strength): two different altitudes (2000m and 3700m), observations at zero, medium and large zenith angles ($0^\circ$, $20^\circ$ and $50^\circ$) and partial moon light. The main conclusions (as detailed in \cite{APP_CTA_MC}) were the following: Several of the proposed layouts could satisfy CTA scientific expectations for most physics cases. In addition, the improvement at the lowest energies ($E < 100$ GeV) of high-altitude sites does not compensate the effective area loss above those energies, favouring sites at moderate altitudes.

The second large-scale MC production (Prod-2) intended to go one step further, with clear objectives: to assess the impact of the construction site on the CTA performance and to search for a feasible layout candidate, taking into account updated cost estimates and new proposed telescope types. With this in mind, telescope responses, site properties and the layout composition were configured with more realistic parameters: 

\paragraph{Telescope response:} Using all \tmtexttt{sim\_telarray} improvements and the updates from all the telescope designs (such as Photo-Multiplier Tubes efficiencies, mirror reflectivity or readout systems), full ray tracing was performed, storing long readout windows per pixel to allow testing of custom trace integration schemes in the analysis and including 3 different trigger schemes: majority, analog-sum and digital-sum.

\paragraph{Layout composition:} New large macro-layouts of telescopes were defined (one for each simulated hemisphere) with new telescope positions and types: 229 telescope positions in the southern site with up to seven telescope types were simulated (LSTs, DC-MSTs, SC-MSTs, and up to four variants of SSTs). For northern sites three telescope types (LSTs, DC-MSTs and SSTs) were simulated at a total of 61 positions.

\paragraph{Site properties:} For each candidate site, specific atmospheric density profiles, altitudes and geo-magnetic fields (direction and intensity) are used. A total of 3 Northern and 6 Southern Hemisphere sites were simulated.

\paragraph{}  Simulated showers include gammas (both point source and diffuse) and background (protons, nuclei, and electrons), with protons ($\sim$ 100 billion events per site) being the particle type consuming most of the CPU time and disk space resources, even though few of them trigger and pass the gamma selection cuts. Most of the simulations are produced for zenith angles of 20$^\circ$, with showers coming from North and from South to include the impact of the geomagnetic field. In addition, dedicated productions were devoted to increased levels of Night Sky Background or to different telescope separations. The Prod-2 simulations were carried out on the CTA computing Grid as well as on the MPIK and  DESY computer clusters.

\section{Analysis}
\label{sec:Analysis}

To evaluate the performance of a telescope layout, analysis tools similar to the ones used in the current IACT experiments are employed. Results presented here use methods based on the MAGIC Analysis and Reconstruction Software (MARS) \cite{MARS}, the software package used by the MAGIC collaboration. The analysis performs the following steps:

\begin{itemize}

\item Trace integration: Pixel charge and the time of arrival are extracted from \tmtexttt{sim\_telarray} output for each triggered event. This is performed in a 2-step process, in which a fit to the photon arrival times for the most significant pixels is subsequently used to search for dimmer signals in a reduced time window. 

\item Image cleaning and parameterization: A traditional Hillas-like parameterization of the images is then performed. 

\item Stereo reconstruction: Using the information gathered by all triggered telescopes observing an event, a stereoscopic reconstruction is performed that measures the energy and direction of the primary gamma ray. Multivariate event classification algorithms are used for background suppression, discerning between cosmic and $\gamma$-ray initiated showers.

\item Performance estimation: Optimal cuts are determined in order to calculate the resulting performance, expressed by the Instrument Response Functions (IRFs). Using the shower's arrival direction and telescope multiplicity, differential sensitivity is maximized for each energy bin. Sensitivity is computed requiring five standard deviations (5 $\sigma$) at each energy bin (equation 17 from \cite{LiMa} considering a ratio of background to signal exposure of 5). In addition, the excess needs to be larger than 10 $\gamma$-rays, and must correspond to a rate higher than the 5\% of the residual background rate.

\end{itemize}


\begin{minipage}[!h]{0.45\textwidth}

It is worth to note that alternative analysis tools have been developed for the analysis of CTA MC data, based on different software packages belonging to different IACT experiments. Fig. \ref{fig:analysis_compare_2Q}\color{black}~ shows the differential sensitivity of the ``2Q" layout in 50 hours calculated with 3 different analysis chains (all introduced in \cite{APP_CTA_MC}): The \textit{Baseline} analysis at the MPIK, the \textit{evndisplay} based analysis at DESY and the MARS based one detailed here. Although some differences appear, results are consistent taking into account the differences between analyses (image cleaning, shower reconstruction, quality cuts and background rejection power).    

\end{minipage}%
 \hspace{0.02\textwidth}
 \begin{minipage}[!h]{0.50\textwidth}
\centering
  \includegraphics[width=0.9\textwidth]{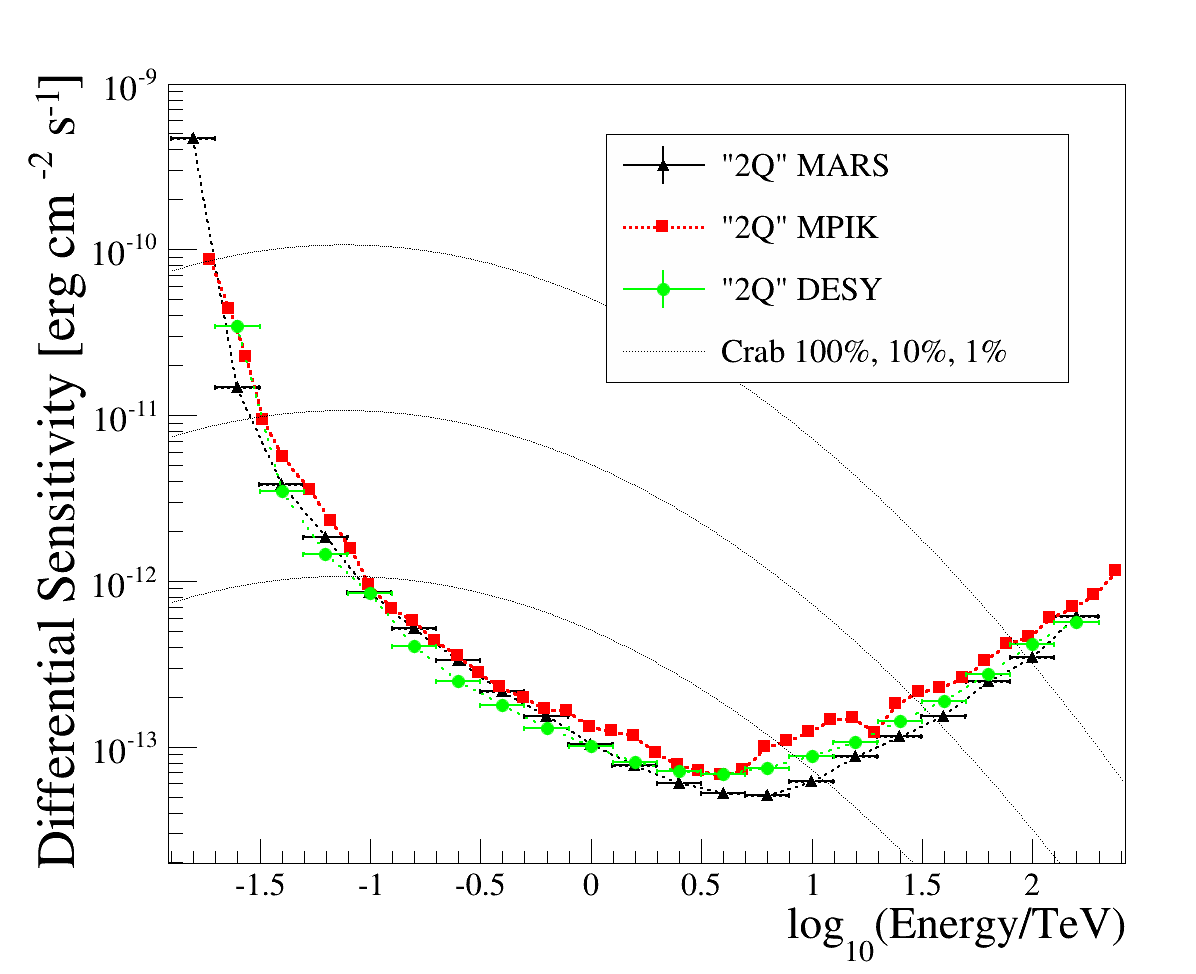}
\captionof{figure}{Differential sensitivity for the CTA-S ``2Q" candidate array (50 hours of observation, N/S pointing average) calculated with 3 alternative analysis chains: \textit{Black}: MARS analysis at IFAE. \textit{Red}: \textit{Baseline} analysis at MPIK. \textit{Green}: \textit{Evndisplay} analysis at DESY.}
\label{trigger_rate_measured}
    \label{fig:analysis_compare_2Q}
    \vspace{0.5 cm}
\end{minipage}


\section{CTA performance}
\label{sec:CTAperformance}

\begin{figure}[!h]
   \centering
    \includegraphics[width=0.7\textwidth]{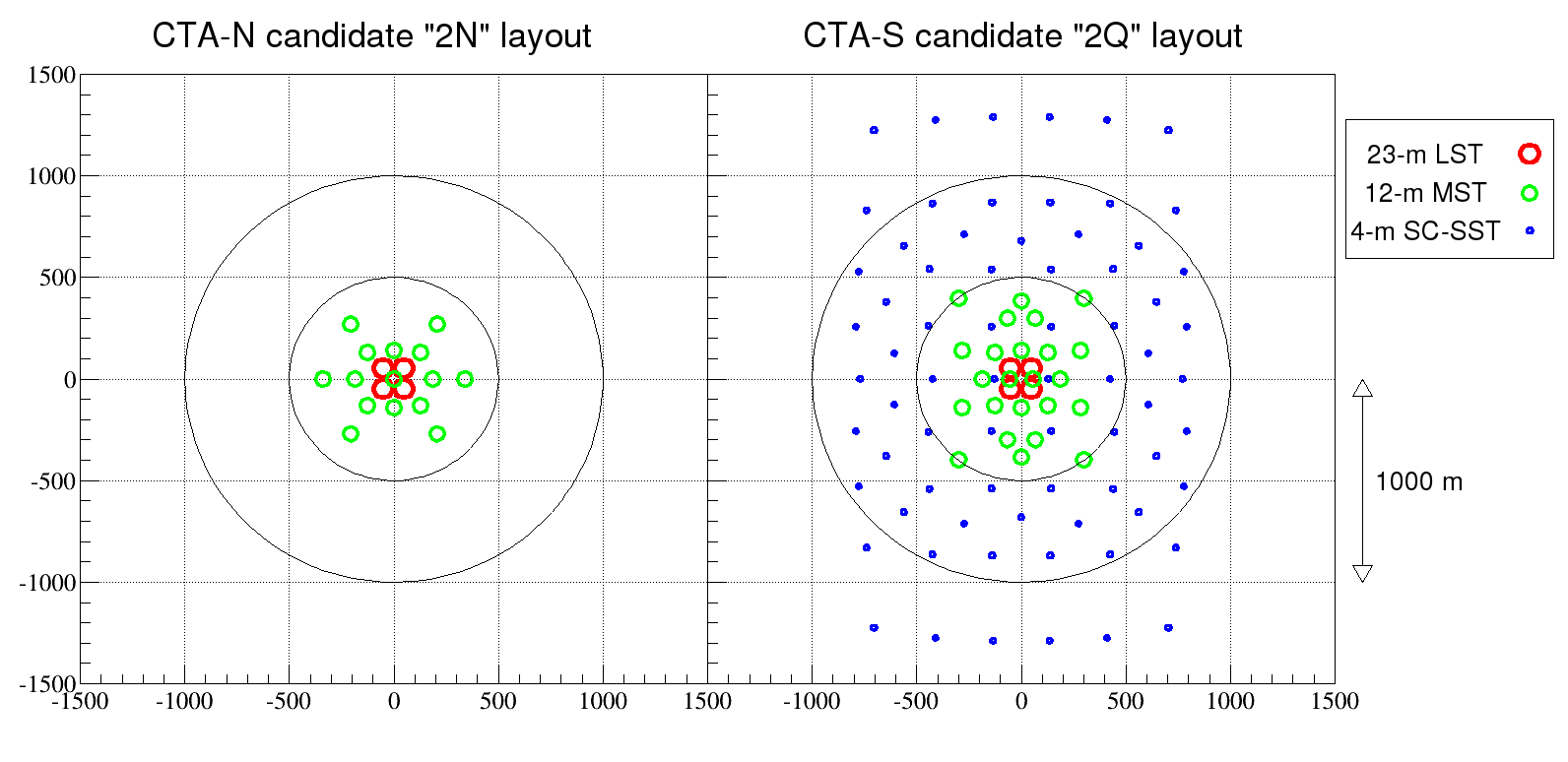}
    \caption{Proposed layouts for the Northern and Southern Hemisphere sites. \textit{Left}: CTA-N candidate layout ``2N", made up of 4 LSTs and  15 MSTs. \textit{Right}: CTA-S candidate layout ``2Q" composed of 4 LSTs, 24 MSTs and 72 4m Schwarzschild Couder (SC) SSTs.}
    \label{fig:2Q_2N_layouts}
\end{figure}

This work focuses on describing the capabilities of two feasible CTA-N and CTA-S layouts. Here, the layouts considered will be the ``2N" candidate array (4 LSTs and 15 MSTs) for the CTA-N layout simulated at Tenerife and the ``2Q" (4 LSTs, 24 MSTs and 72 double mirror SC-SSTs) for the CTA-S simulated at Namibia (Aar), both shown in Fig. \ref{fig:2Q_2N_layouts}\color{black}. The choice of sites is not related in any sense with the final construction location, and is based on availability of the simulated data files and the number of cross-checks with alternative analysis chains. 

Each CTA layout performance is calculated from the average of two simulated observations: north and south pointing directions at 20$^\circ$ of zenith angle. This is done to account for the performance differences coming from the geomagnetic field (GF) effect, detailed in \cite{MC_ICRC_site:2015}. It should be noted that the simulated observation directions for the chosen Southern Hemisphere site are particularly optimistic, as the perpendicular component of the geomagnetic field for the north pointing direction is close to zero. Differences in the differential sensitivity caused by the GF between CTA-N and CTA-S decrease by $\sim$ 35\% when considering a realistic distribution of observation directions in zenith and azimuth, keeping in mind that Northern Hemisphere sites have, on average, higher perpendicular GF components.

\begin{figure}[!h]
\begin{minipage}[c]{0.6\textwidth}%
\includegraphics[width=\textwidth]{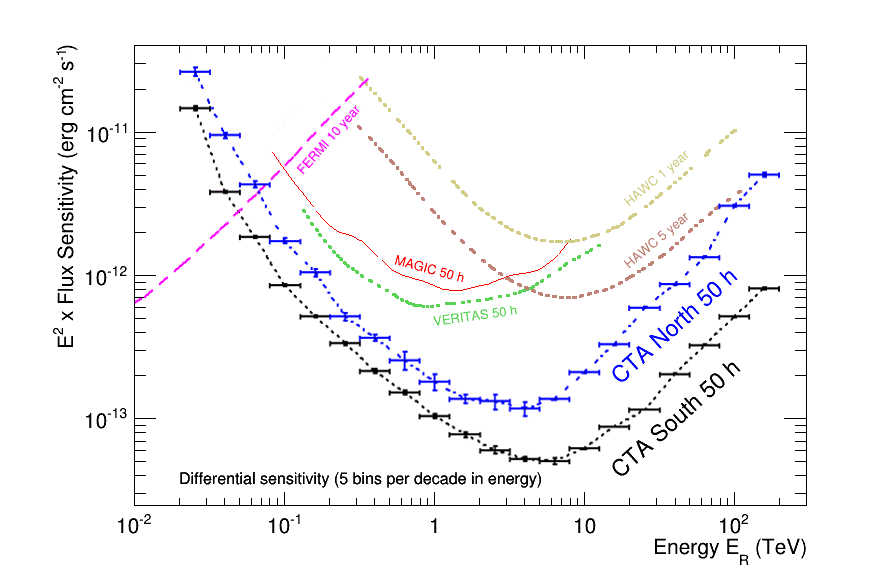}
\end{minipage}\hfill
\begin{minipage}[c]{0.4\textwidth}%
\caption{Differential sensitivity for a point-like $\gamma$-ray source of the CTA-N and CTA-S candidate arrays (50 hours of observation, N/S pointing average) together with the current MAGIC \cite{MAGIC:performance} and VERITAS (50 hours) and and future \textit{Fermi}-LAT \cite{Fermi_vs_CTA} (over 10 years of operation) and HAWC \cite{HAWC:performance} (1 and 5 years) attained sensitivities.}
\label{fig:diffsens_comparison}
\end{minipage}
\end{figure}

The minimum detectable flux in 50 hours from a steady $\gamma$-ray point-like source as a function of the energy (differential sensitivity) for both CTA site candidates is shown in Fig. \ref{fig:diffsens_comparison}\color{black}, together with the current MAGIC \cite{MAGIC:performance} and VERITAS\footnote{\url{http://veritas.sao.arizona.edu/}} (during 50 hours) and future HAWC \cite{HAWC:performance} and \textit{Fermi}-LAT \cite{Fermi_LAT:performance} sensitivities. Both sites attain good performance over four decades of energy, and would be able to measure a spectrum 100 times fainter than the Crab Nebula (10 mCrab) across an order of magnitude in energy in 10 hours of observation for the case of CTA-S, increasing to 25 hours for CTA-N. 


\begin{figure}[!h]  
  \centering
  \includegraphics[width=0.49\textwidth]{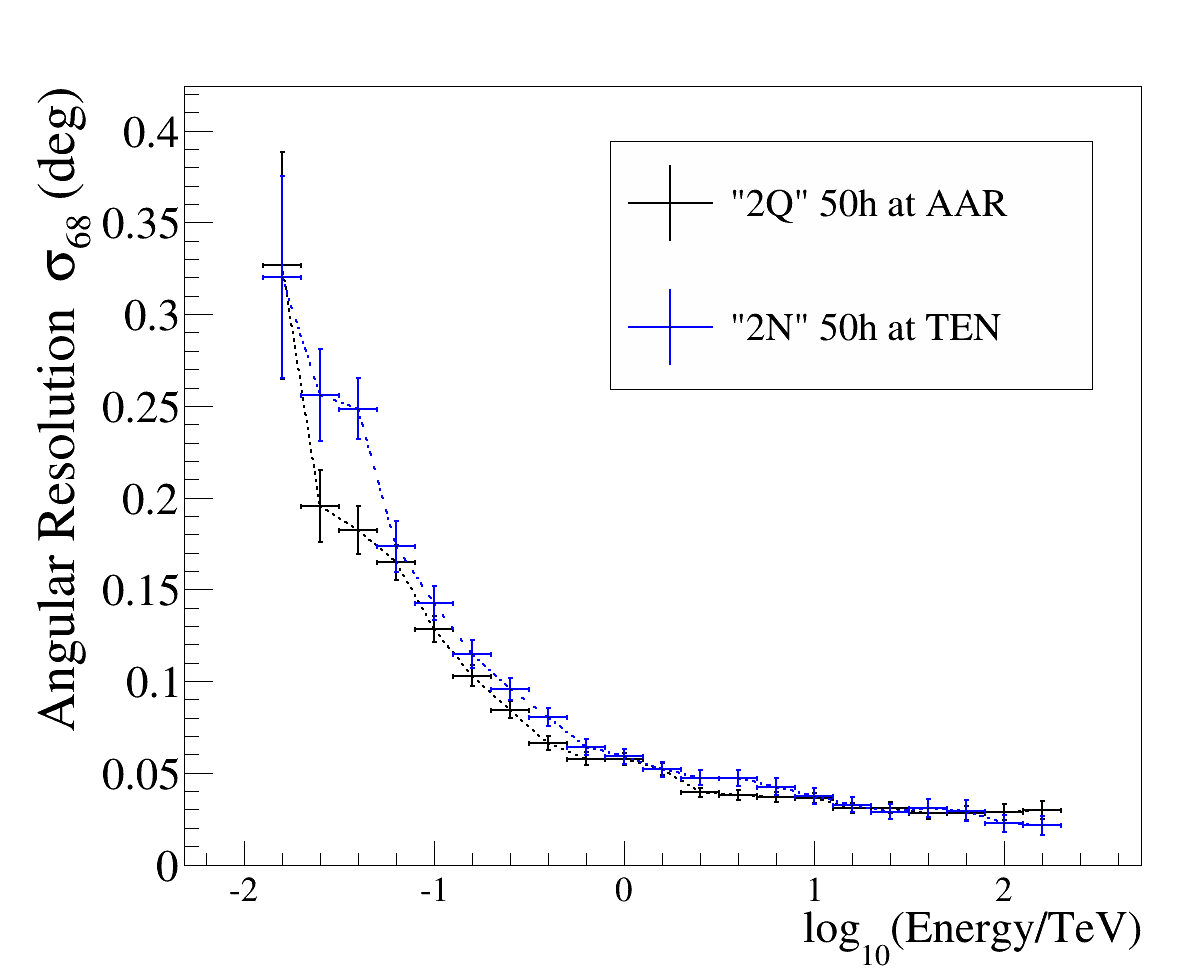} \includegraphics[width=0.49\textwidth]{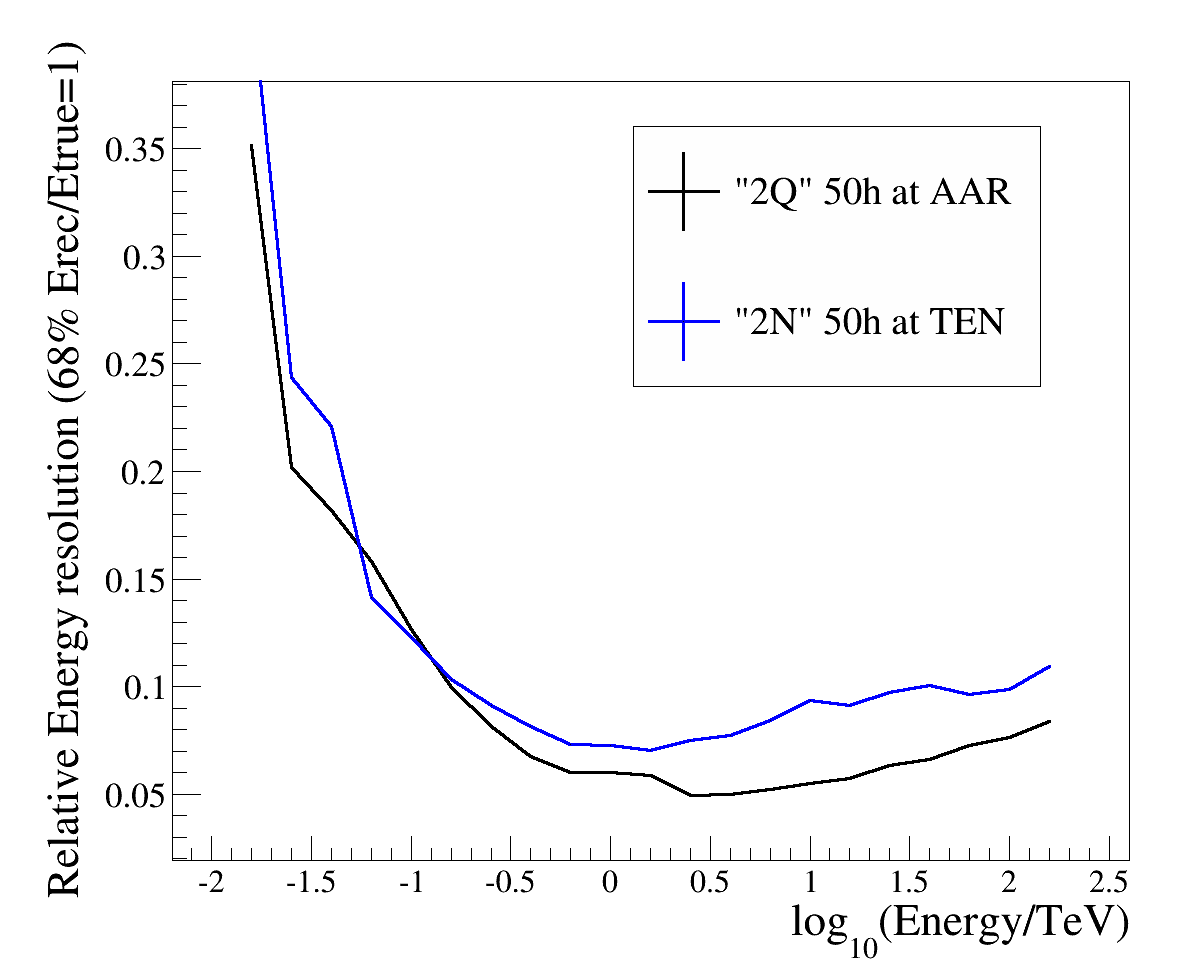}
  \caption{Angular and energy resolution for both CTA-N and CTA-S candidates with respect to the reconstructed energy: (\textit{Left}) The angular resolution, defined as the angle containing the 68\% of the reconstructed gamma-rays, relative to the true direction. (\textit{Right}) The energy resolution is defined such that 68\% of gamma rays will have true energy within $\Delta$E of their reconstructed energy.}
  \label{fig:ang_e_res}
\end{figure}

As shown in Fig. \ref{fig:ang_e_res}\color{black}, the CTA Observatory will not only provide an improvement in sensitivity with respect to the current instruments; it will also improve the angular and energy resolution, shrinking by a factor $\sim$ 3 the radius of the Point Spread Function (PSF) with respect to current instruments, resolving structures on the order of 2 arc minutes at the highest energies observed by CTA, becoming the $\gamma$-ray detector with the best angular resolution ever constructed. Energy resolution will also improve significantly, reaching values close to the 5\% in fractional uncertainty. 

To understand the overall performance of such a system, it is required to recognize the contribution of each telescope type at different energy ranges. Performing independent analyses of sub-layouts of each telescope type within the CTA-S candidate array, it is possible to see the impact of each telescope on the differential sensitivity. Fig. \ref{fig:2Q_subtelescopes}\color{black}~ shows the detached performance of each layout of individual telescope types contained in the CTA-S candidate: 


\begin{figure}[!h]
\begin{minipage}[c]{0.6\textwidth}%
\includegraphics[width=\textwidth]{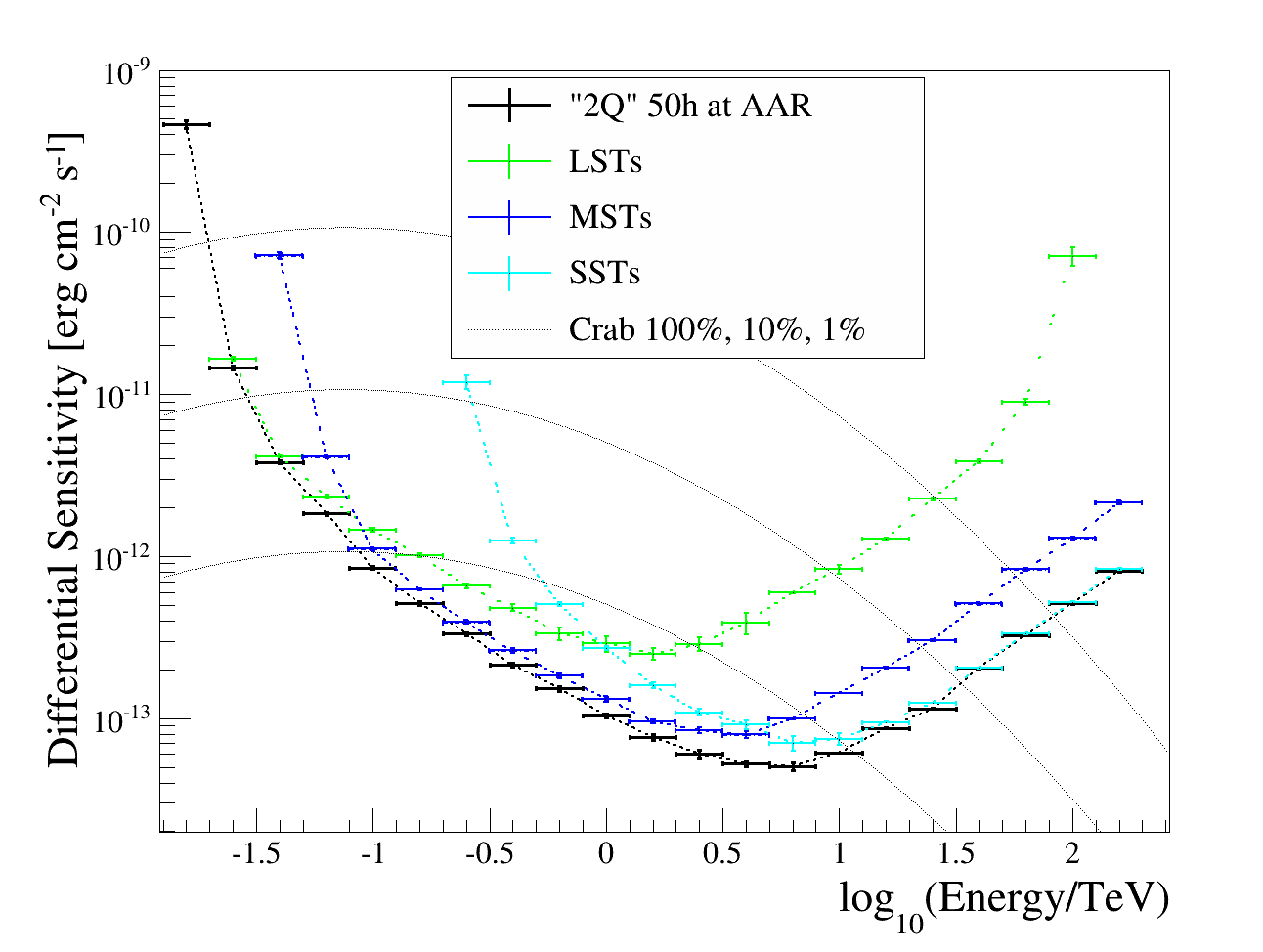}
\end{minipage}\hfill
\begin{minipage}[c]{0.4\textwidth}%
\caption{Different telescope type contributions to the over-all differential sensitivity of CTA-S candidate layout ``2Q", simulated at the Namibian site (50 hours, north/south pointing average). LSTs govern the low energy range below 100 GeV. MSTs dominate in the CTA core energies, up to 5 TeV, where SSTs start to monopolize performance.}\label{fig:2Q_subtelescopes}
\end{minipage}

\end{figure}

\begin{itemize}
\item \textbf{Below 100 GeV}: The subset of 4 LSTs dominates the sensitivity below 100 GeV. Although this range may seem small, a large number of physics cases will benefit from the CTA performance in this region, and cross-calibration with other $\gamma$-ray detectors such as \textit{Fermi}-LAT is desired.

\item \textbf{From 100 Gev to 5 TeV}: The layout of 24 MSTs dominates the instrument performance in the core energy range, above 100 GeV up to 5 TeV. Their contribution to the lower energies is also significant, as they allow to reconstruct events of $\sim$ 60 GeV (boosting sensitivity by $\sim$ 20\%). They also improve gamma-hadron separation of events observed by the LSTs, even at the lowest energies, as they detect possible $\pi_{0}$ initiated electromagnetic sub-showers or muons originated in hadronic EASs. They also contribute to higher energies up to $\sim$ 10 TeV.

\item \textbf{Above 5 TeV}: The array of 72 SC-SSTs, covering an area of $\sim$ 4 km$^2$, dominates the sensitivity above 5 TeV (CTA-S only). Comparison of alternative layouts included in the Prod-2 simulations shows that the performance of about 70 SC-SSTs (of 4 m diameter) is comparable to that of about 35 DC-SSTs of 7 m diameter when both arrays cover a similar area.

\end{itemize}

Off-axis capabilities are also crucial for a significant number of the CTA key science projects. Sky surveys and studies of diffuse and extended sources of gamma-ray emission will benefit from a wide field of view (FoV) and accurate event reconstruction away from the camera center. To characterize the observatory performance for different off-axis angles, the differential sensitivity of a point-like source located at different distances from the center of the camera is estimated. 

Below 100 GeV, performance is dominated by the LSTs, which are telescopes with a smaller FoV. Their sensitivity does not decrease significantly at distances from the camera center smaller than 2$^{\circ}$, but declines sharply at larger distances (by a factor 4 at a 2.5$^{\circ}$). For the MSTs, their modified Davies-Cotton optics improve off-axis performance with respect to the parabolic one. Between 100 GeV to 5 TeV, the sensitivity drops by less than a factor 2 on sources 3$^{\circ}$ away of the center of the camera. Above 5 TeV the off-axis performance improves significantly. This effect is related to the good off-axis performance of the SC-SSTs dominating at these energies. Their performance is expected to be approximately flat up to 5$^{\circ}$ away of the camera.

\section{Conclusions}

The analysis presented in this work of both feasible CTA-N and CTA-S candidates represent a detailed estimation of the future CTA capabilities, capable of fulfilling all scientific goals defined by the Consortium. With the results from both CTA large-scale MC productions, layouts have been optimized to find the best trade-off between event quantity and reconstruction quality. The search for the most efficient layout for each site is still on-going, and will likely be completed in the third large-scale MC production (Prod-3), which is currently in preparation. A denser layout of 4m-SSTs is capable of attaining similar performance as the one initially proposed with 7m-SSTs. In addition, the newly designed Schwarzschild-Couder SSTs show great off-axis capabilities providing, together with the MSTs, a FoV with a radius larger than 3$^\circ$ for the whole energy range above 100 GeV, increasing up to $\sim$ 5$^\circ$ at the highest energies. 

MC simulations provide a crucial test-bench for the different designs within the CTA project, and Prod-2 results demonstrate that their correct implementation would attain the desired performance and potential scientific output.

\begin{acknowledgments}
\small
This work is partially funded by the ERDF under the Spanish MINECO grant FPA2012-39502 and FPA2010-22056-C06-06. We gratefully acknowledge support from the agencies and organizations  listed under Funding Agencies at this website: http://www.cta-observatory.org/.

\end{acknowledgments}


\bibliographystyle{JHEP}

\bibliography{skeleton} 

\end{document}